\newif\ifapsclass
\IfFileExists{revtex4-2.cls}{\apsclasstrue\documentclass[
aps,
prl,
twocolumn,
superscriptaddress,
nofootinbib,
floatfix
]{revtex4-2}}{\apsclassfalse\documentclass[10pt,twocolumn]{article}\usepackage[margin=0.68in,columnsep=0.24in]{geometry}}

\usepackage{amsmath,amssymb,bm}
\usepackage{graphicx}
\usepackage{xcolor}
\usepackage{microtype}
\usepackage[colorlinks=true,citecolor=blue,linkcolor=blue,urlcolor=blue]{hyperref}

\graphicspath{{figures/}}

\newcommand{\ii}{\mathrm{i}}
\newcommand{\hc}{\mathrm{H.c.}}
\newcommand{\MBZ}{\mathrm{MBZ}}
\newcommand{\C}{\mathcal C}
\newcommand{\tp}{t_\perp}

\begin{document}

\title{
Controlling topology in flux-mismatched Hofstadter bilayers}

\ifapsclass
\author{Adel Ali}
\author{Alexey Belyanin}
\affiliation{Department of Physics and Astronomy, Texas A\&M University, College Station, Texas 77843, USA}
\else
\author{Adel Ali and Alexey Belyanin\\
\normalsize Department of Physics and Astronomy, Texas A\&M University, College Station, Texas 77843, USA}
\fi

\date{\today}

\newcommand{\paperabstract}
{Stacked two-dimensional materials provide a promising platform for electrically controlling topological electronic states. However, tunneling between layers hybridizes their bands and removes the crossings needed to change topology.  
We show that this does not always have to be the case.  Band crossings and associated Weyl points are topologically enforced in Hofstadter bilayers whose layers experience different magnetic fluxes. When resonant magnetic Bloch multiplets carry unequal Chern numbers, their projected tunneling is topologically obstructed and must vanish at isolated momenta. Sweeping the layer bias through these zeros creates synthetic Weyl monopoles in momentum–bias space that transfer the Chern mismatch. We demonstrate two consequences: a direct transition between insulating Chern phases and a reentrant compensated metal in which Lifshitz transitions bound the metallic window while internal Weyl events reconstruct the band topology. Consequently, the fixed-filling Hall response remains continuous and nonquantized even as integer Chern number is transferred between bands. Berry-flux, TKNN, and interface calculations independently verify the mechanism and its multichannel chiral signature. We outline realizations in Moiré and anomalous-Hall heterostructures, establishing flux mismatch as an experimentally accessible route to electrically programmable Chern phases and chiral transport.}

\ifapsclass
\begin{abstract}\paperabstract\end{abstract}
\maketitle
\else
\maketitle
\begin{abstract}\paperabstract\end{abstract}
\fi

\textit{Introduction.---}
The Hofstadter problem elegantly combines magnetic translation symmetry, fractal spectra, and quantized Hall topology \cite{Hofstadter1976}. However, its implementation usually requires unrealistically high magnetic fields. Moir\'e superlattices created by twisting adjacent monolayers of 2D materials have made appreciable flux per unit cell experimentally accessible and have exposed Hofstadter minibands and correlated Chern phases~\cite{Dean2013,Hunt2013,Ponomarenko2013,Spanton2018,Chen2020Chern}.  They also raise an appealing  possibility of electrically tunable topological phases \cite{Ding2025,Perrin2024,Patra2023}. However, there is a fundamental impediment: interlayer tunneling ordinarily gaps a bias-driven band inversion needed to change band topology. Here we propose a solution to this problem. Hofstadter systems normally involve one magnetic translation algebra.  Here we ask what happens  when local tunneling is forced to hybridize two \emph{different} magnetic translation algebras.

\begin{figure*}[t]
\centering
\includegraphics[width=\textwidth]{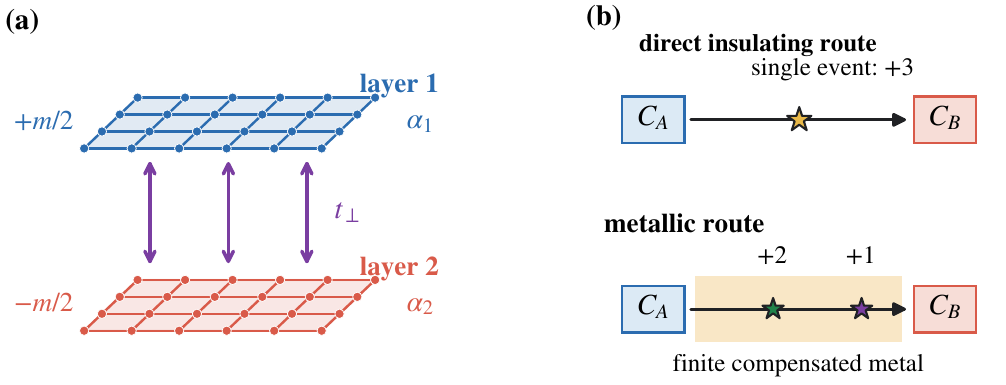}
\caption{\textbf{Obstructed tunneling and two spectral routes for Chern transfer.}
(a) Two locally coupled layers experience fluxes $\alpha_1\ne\alpha_2$ and opposite shifts $\pm m/2$.
(b) The same endpoint transfer can occur via a ``clean''  route, i.e., through an isolated single insulating critical bias or through a finite compensated metal.   Topology fixes the net transferred charge, whereas energetics fixes the route. Here $C_B - C_A = 3$ corresponding to the example in the text.}
\label{fig:concept}
\end{figure*}

We consider two layers with fluxes $\alpha_1\ne\alpha_2$, which could be realized, e.g., when two adjacent Moir\'e superlattices have different unit cell areas.  A perpendicular displacement field, or an internal-state detuning in a synthetic implementation, shifts the two Hofstadter spectra through one another. Interlayer tunneling normally converts a band crossing into an avoided crossing. We show that it cannot do so everywhere when the two resonant Bloch multiplets have unequal Chern numbers. When bands with unequal Chern numbers invert, their hybridization cannot remain nonzero throughout the magnetic Brillouin zone: topology forces zeros of the projected tunneling.  A bias sweep through these zeros produces Weyl points in the three-dimensional parameter space $(k_x,k_y,m)$, where $m$ is a band shift due to  displacement field, and transfers their Berry charge between the hybridized bands.  Related overlap nodes and coherence vortices occur whenever states with different Chern indices are compared~\cite{Huang2016,Bultinck2020}; here they become a single-particle control mechanism for a Hall band structure. 
Figure~\ref{fig:concept} summarizes the proposal.  A local bias can select insulating domains with Chern numbers $C_A$ and $C_B$.  Their interface carries $|C_B-C_A|$ net chiral modes. The essential control parameter is electrical; the flux pattern can remain fixed.


This mechanism has an especially revealing consequence in a metal.  Noninteracting insulator--semimetal--insulator sequences driven by an electric field are known~\cite{Perrin2024,Patra2023}, as are direct-gap Chern-metal regimes~\cite{Hu2015ChernMetal,Sorn2018}.  Here the finite compensated-metal interval contains a \emph{topologically enforced cascade}: electron and hole pockets first appear at a Lifshitz boundary, subsequent Weyl events transfer 
$|C_B-C_A|$ Chern units while the system remains metallic, and a final Lifshitz boundary restores an indirect gap. Away from the isolated Weyl biases, the relevant band remains separated from the next band at every fixed momentum, $\Delta_{\rm dir}>0$, even though their energy ranges overlap, $\Delta_{\rm ind}<0$. Its Bloch bundle and Chern number therefore remain well defined despite the absence of a fixed-filling insulating gap. Thus the Fermi surface and the topology of this directly isolated band reconstruct at different control parameters.

\textit{Model and Chern-transfer theorem.---}
On a square lattice in the Landau gauge, we consider the Hamiltonian 
\begin{align}
H={}&-t\sum_{x,y,\ell}\left(c^\dagger_{x+1,y,\ell}c_{x,y,\ell}+\hc\right)
\nonumber\\
&-t\sum_{x,y,\ell}\left(e^{\ii2\pi\alpha_\ell x}
c^\dagger_{x,y+1,\ell}c_{x,y,\ell}+\hc\right)\nonumber\\
&+\frac{m}{2}\sum_{x,y}(n_{x,y,1}-n_{x,y,2})\nonumber\\
&+\tp\sum_{x,y}(c^\dagger_{x,y,1}c_{x,y,2}+\hc).
\label{eq:realspace}
\end{align}
For $\alpha_i=p_i/q_i$, the common magnetic period is $Q=\operatorname{lcm}(q_1,q_2)$.  Both momenta are $2\pi$ periodic when $k_x$ is the phase acquired across the $Q$-site cell.  The Bloch Hamiltonian is
\begin{equation}
\mathcal H(\bm k;m)=
\begin{pmatrix}
H_{\alpha_1}(\bm k)+mI_Q/2&\tp I_Q\\
\tp I_Q&H_{\alpha_2}(\bm k)-mI_Q/2
\end{pmatrix},
\label{eq:bloch}
\end{equation}
where $[H_\alpha]_{xx}=-2t\cos(k_y+2\pi\alpha x)$, nearest-neighbor $x$ hopping is $-t$, and the cell-boundary hopping is $-t e^{\ii k_x}$~\cite{Hofstadter1976}.

For unequal denominators, a clean endpoint label requires complete folded multiplets at both biases.  Writing $g=\gcd(q_1,q_2)$, the smallest common rank and the numbers of original Hofstadter bands are
\begin{equation}
n_{\rm occ}^{\min}=\frac{Q}{g},\qquad r_i=\frac{q_i}{g}.
\label{eq:compatible}
\end{equation}
In our convention their endpoint Chern numbers follow from the TKNN congruence~\cite{TKNN1982,Wannier1978,DanaAvronZak1985}
\begin{equation}
p_i C_i\equiv-r_i\pmod{q_i},\qquad
\Delta C=C_2-C_1,
\label{eq:tknn}
\end{equation}
with the centered integer solution chosen.  At the minimal compatible rank this gives the explicit design rule
\begin{equation}
C_i=\operatorname{center}_{q_i}\!\left[-\frac{q_i}{g}p_i^{-1}\right],
\qquad \Delta C=C_2-C_1,
\label{eq:endpoint-rule}
\end{equation}
where $p_i^{-1}$ is the modular inverse modulo $q_i$.  Thus the flux-mismatch magnitude $|\alpha_2-\alpha_1|$ alone does not determine $\Delta C$; both rational fluxes and the occupied gaps do.  Equation~\eqref{eq:endpoint-rule} fixes the endpoint transfer without diagonalizing the coupled bilayer, but it does not determine how that transfer is distributed among intermediate nodes.

Project now onto two isolated, equal-rank resonant multiplets $A$ and $B$:
\begin{equation}
H_{\rm eff}=
\begin{pmatrix}
\varepsilon_A(\bm k)+m/2&T(\bm k)\\
T^\dagger(\bm k)&\varepsilon_B(\bm k)-m/2
\end{pmatrix}.
\label{eq:projected}
\end{equation}
For line bands, $T=\tp\langle u_A|u_B\rangle$ is the local matrix of a bundle map $E_B\to E_A$.  For rank-$r$ multiplets, its determinant is a section of $\det E_A\otimes(\det E_B)^*$.  Defining $w_j$ as the phase winding of $\det T$ in this convention, the signed zeros obey
\begin{equation}
{\sum_j w_j=C_A-C_B.}
\label{eq:vorticity}
\end{equation}
This is the central obstruction: if $C_A\ne C_B$, local tunneling cannot be invertible everywhere.  Reversing the map, or equivalently using $T^\dagger$, reverses all $w_j$ but leaves the forced number of zeros unchanged.  The scalar and determinant-line-bundle proofs, including the folded-multiplet construction, are given in the Supplemental Material~\cite{SM}.

The same projection makes the transition locations predictive.  For purely interlayer tunneling and multiplets separated from remote bands by $\Delta_{\rm iso}$,
\begin{align}
T(\bm k)&=P_A(\bm k)(\tp\tau_x)P_B(\bm k)
+O(\tp^3/\Delta_{\rm iso}^2),\nonumber\\
m_{c,j}&=\varepsilon_B(\bm k_j)-\varepsilon_A(\bm k_j)
+O(\tp^2/\Delta_{\rm iso}).
\label{eq:predictive}
\end{align}
Here $P_{A,B}(\bm k)$ are the spectral projectors onto the two isolated magnetic Bloch multiplets brought into resonance, and $\tau_x$ is the Pauli matrix in layer space that implements interlayer tunneling.
There is no second-order correction to the off-diagonal map because each tunneling insertion flips the layer; second order only shifts the diagonal energies.  Hence the decoupled-layer overlap predicts the vortex momenta more accurately than the leading energy difference predicts the critical biases.

Consider as an example $\alpha_1 = 1/5$  and $\alpha_2 = 2/5$. In this case the decoupled-layer overlap has zeros at $(k_x,k_y)=(\pi,\pm 0.9052)$ and $(\pi,\pi)$, compared with the exact finite-tunneling nodes $(\pi,\pm 0.9107)$ and $(\pi,\pi)$ at $\tp/t=0.2$.  The leading energy differences predict $m_c/t=0.473$ and $0.569$; including the second-order diagonal shifts gives $0.541$ and $0.684$, compared with the exact values $0.536$ and $0.675$.  Thus the practical decoupled-layer prediction is accurate to $0.9\%$ and $1.4\%$, respectively.  A tunneling-strength sweep verifies the predicted $O(\tp^2)$ displacement of the overlap zero and the $O(\tp^4)$ residual of the second-order resonance formula when evaluated at the corrected zero~\cite{SM}. Figure~\ref{fig:metal}(d) compares the resulting critical-bias prediction
with exact full-bilayer calculations over a tunneling-strength sweep for
both flagship examples.  The decoupled-layer overlap together with the
second-order diagonal correction tracks the exact transition biases and
remains within $1.5\%$ at $\tp/t=0.2$.
Near each simple zero the projected Hamiltonian is a Weyl cone in
$(k_x,k_y,m)$. Whenever the lowest $n_{\rm occ}$  bands are directly
separated from the remaining bands, we define the Chern number of this
rank-$n_{\rm occ}$ Bloch bundle as
\begin{equation}
\C(m)=\frac{1}{2\pi}
\int_{\MBZ} d^2k\,
\operatorname{Tr}\mathcal F_{xy}^{\rm occ}(\bm k;m),
\label{eq:bundle-chern}
\end{equation}
where $\mathcal F_{xy}^{\rm occ}$ is the non-Abelian Berry curvature of
the lowest $n_{\rm occ}$ eigenstates at each momentum.  Importantly,
$\C(m)$ remains well defined when $\Delta_{\rm ind}<0$ as long as
$\Delta_{\rm dir}>0$: an indirect energy overlap makes the fixed-filling
system metallic but does not destroy the momentum-resolved Bloch bundle.
The invariant becomes undefined only at a direct-gap closing.

The closed-surface Berry flux $\chi_j$ of a Weyl node changes this bundle
Chern number according to
\begin{equation}
\C(m_2)-\C(m_1)
=\sum_{m_1<m_{c,j}<m_2}\chi_j,
\quad
\sum_j\chi_j=C_B-C_A.
\label{eq:weyltransfer}
\end{equation}
With the winding convention of Eq.~\eqref{eq:vorticity}, a simple node
has $\chi_j=-w_j$; this sign converts the determinant-map index into the
physical endpoint transfer.
 With the winding convention of Eq.~\eqref{eq:vorticity}, a simple node has $\chi_j=-w_j$; this sign converts the determinant-map index into the physical endpoint transfer.  The model also has the antiunitary symmetry
\begin{equation}
\Theta=R_xK,\quad
\Theta\mathcal H(k_x,k_y;m)\Theta^{-1}
=\mathcal H(k_x,-k_y;m),
\label{eq:pairing-symmetry}
\end{equation}
where $R_x$ reflects $x\mapsto-x$.  Nodes away from $k_y=0,\pi$ are therefore paired at opposite $k_y$ with equal monopole charge, whereas a unit-charge node on either fixed line may occur alone.  This symmetry explains the $+2$ followed by $+1$ pattern below, although it does not by itself pin the nodes to $k_x=\pi$.

We compute $\C$ using the gauge-invariant lattice construction~\cite{Fukui2005}.  We distinguish
\begin{align}
\Delta_{\rm dir}&=\min_{\bm k}[E_{n_{\rm occ}+1}(\bm k)-E_{n_{\rm occ}}(\bm k)],\nonumber\\
\Delta_{\rm ind}&=\min_{\bm k}E_{n_{\rm occ}+1}(\bm k)
-\max_{\bm k}E_{n_{\rm occ}}(\bm k). \nonumber 
\end{align}
A positive direct gap protects the band bundle; a positive indirect gap is additionally required for a fixed-filling insulator and quantized Hall plateau.

\begin{figure}[t]
\centering
\includegraphics[width=\linewidth]{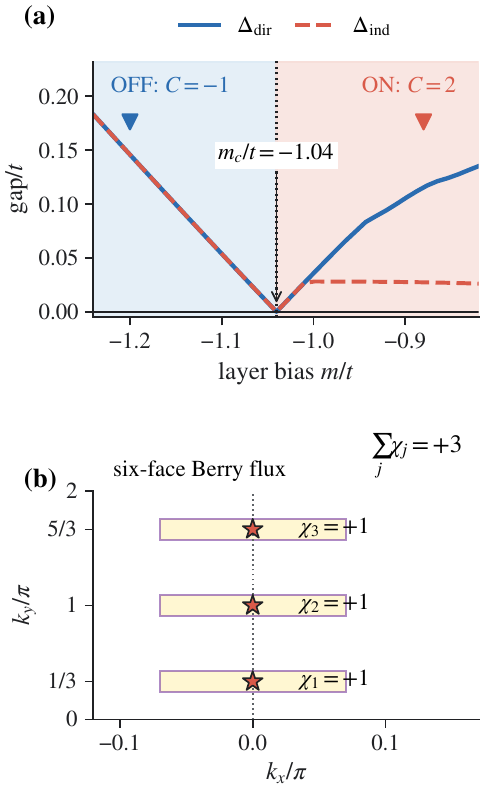}
\caption{\textbf{Direct insulating $C=-1\to2$ transfer.}
(a) Direct and indirect gaps for
$(\alpha_1,\alpha_2)=(1/3,5/6)$,
$n_{\rm occ}=2$, and $\tp/t=0.2$.
Both gaps vanish at the common critical bias $m_c/t=-1.04$ and reopen
immediately on either side.
(b) The three nodes at $k_x=0$ are independently enclosed by small
surfaces in $(k_x,k_y,m)$; integration of the occupied-frame Berry flux
through all six faces gives $\chi_j=+1$ for each node and
$\sum_j\chi_j=3$.}
\label{fig:clean}
\end{figure}

\textit{Clean (direct insulating) route.---}
As an example, consider the different-denominator pair $(\alpha_1,\alpha_2)=(1/3,5/6)$ has $Q=6$, $g=3$, and the minimal compatible rank $n_{\rm occ}=2$.  Equation~\eqref{eq:tknn} predicts $C_A=-1$, $C_B=2$, and $\Delta C=3$.  At $\tp/t=0.2$, the full 12-band Hamiltonian has three symmetry-related degeneracies,
\begin{align}
\frac{m_c}{t}&=-1.04,\nonumber\\
(k_x,k_y)&=\left(0,\frac{\pi}{3}\right),(0,\pi),
\left(0,\frac{5\pi}{3}\right).
\label{eq:cleancritical}
\end{align}
Both direct and indirect gaps close only at this bias and immediately reopen, yielding a direct insulating $C=-1\to2$ transition [Fig.~\ref{fig:clean}(a)].  We independently enclose each degeneracy by a small box in $(k_x,k_y,m)$ and integrate the occupied-frame Berry flux through all six faces.  Every box gives $\chi_j=+1$, stable from $7\times7$ through $25\times25$ meshes per face [Fig.~\ref{fig:clean}(b)].  The total closed-surface charge is therefore $+3$, without inferring it from a difference of constant-$m$ Chern numbers.

The direct insulating transition establishes that the obstruction survives common-cell folding and that the charge can be delivered at a single critical bias.  It does not require an intervening metal.  Conversely, topology does not require the interpolation to remain insulating: dispersion can close the indirect gap before any Weyl event occurs.

\begin{figure}[t]
\centering
\includegraphics[width=\linewidth]{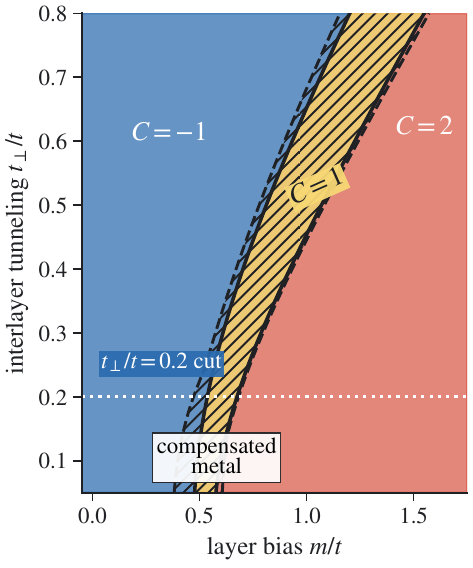}
\caption{\textbf{Finite reentrant-metal region.}
For $(\alpha_1,\alpha_2)=(1/5,2/5)$, solid curves are direct-gap/Weyl transitions and dashed curves are the Lifshitz boundaries $\Delta_{\rm ind}=0$.  Their separation encloses a finite compensated-metal region in the $(m,\tp)$ plane.  The dotted line marks the cut $\tp/t=0.2$ analyzed in Fig.~\ref{fig:metal}. 
}
\label{fig:phase-map}
\end{figure}

\textit{Chern transfer inside a compensated metal.---}
We now turn to an example with the equal-denominator pair $(1/5,2/5)$ at $n_{\rm occ}=1$.  Figure~\ref{fig:phase-map} shows that the reentrant metal occupies a finite region rather than a fine-tuned cut.  At $\tp/t=0.2$, four distinct events occur as we vary bias:
\begin{align}
\frac{m_L}{t}=0.472
&<\frac{m_{c1}}{t}=0.536\nonumber\\
&<\frac{m_{c2}}{t}=0.675
<\frac{m_R}{t}=0.682.
\label{eq:fourbiases}
\end{align}
The indirect gap becomes negative at $m_L$, producing one hole pocket and two electron pockets with exactly compensated total areas at fixed filling.  The pockets occupy at most $0.15\%$ of the magnetic Brillouin zone.  They disappear only at $m_R$ (Fig.~\ref{fig:metal}(a,b)).

Inside this metallic interval, the directly isolated
$n_{\rm occ}=1$ bundle undergoes two topological reconstructions,
\begin{equation}
\C=-1\xrightarrow[\,m_{c1}\,]{+2}1
\xrightarrow[\,m_{c2}\,]{+1}2.
\label{eq:cascade}
\end{equation}
The first transfer is carried by two symmetry-related nodes at $(k_x,k_y)=(\pi,\pm 0.9107)$, while the second occurs at $(\pi,\pi)$.  Independently of the slice-Chern jumps, we enclose these three nodes in separate six-face boxes in $(k_x,k_y,m)$.  The outward Berry flux gives $(\chi_+,\chi_-,\chi_\pi)=(+1,+1,+1)$ on every mesh from $7\times7$ through $25\times25$ vertices per face, directly measuring $+2$ at $m_{c1}$ and $+1$ at $m_{c2}$; all box boundaries remain directly gapped~\cite{SM}.  Equation~\eqref{eq:weyltransfer} fixes their total charge $+3$, but not their critical biases or ordering.  Those are set by energies $\varepsilon_B(\bm k_j)-\varepsilon_A(\bm k_j)$.

The physical fixed-filling Hall conductivity does not jump by these integers.  At every finite temperature the chemical potential is recomputed from the fixed-filling number equation.  At the first node both degenerate states remain predominantly occupied, while at the second they are predominantly empty.  Their singular two-band contributions therefore cancel in the Fermi-sea response, giving the continuous, noninteger values $\sigma_{xy}(m_{c1})\simeq0.05\,e^2/h$ and $\sigma_{xy}(m_{c2})\simeq1.49\,e^2/h$ at $T/t=0.002$ [Fig.~\ref{fig:metal}(c)].  This is not a new symmetry class or a disorder-protected critical metal.
The single-particle Hamiltonian belongs to Altland--Zirnbauer class A:
it has no conventional time-reversal, particle-hole, or chiral symmetry
and in two dimensions admits an integer Chern invariant
~\cite{Schnyder2008,Ryu2010}.  Its unusual feature is instead the
separation between the Lifshitz transitions, which determine when a
Fermi surface exists, and the synthetic Weyl monopoles, which determine
when the directly isolated band changes topology.  The pocket convergence, Kubo calculation, longitudinal response, edge resonance, and localization diagnostics are given in Ref.~\cite{SM}.

\begin{figure*}[t]
\centering
\includegraphics[width=\textwidth]{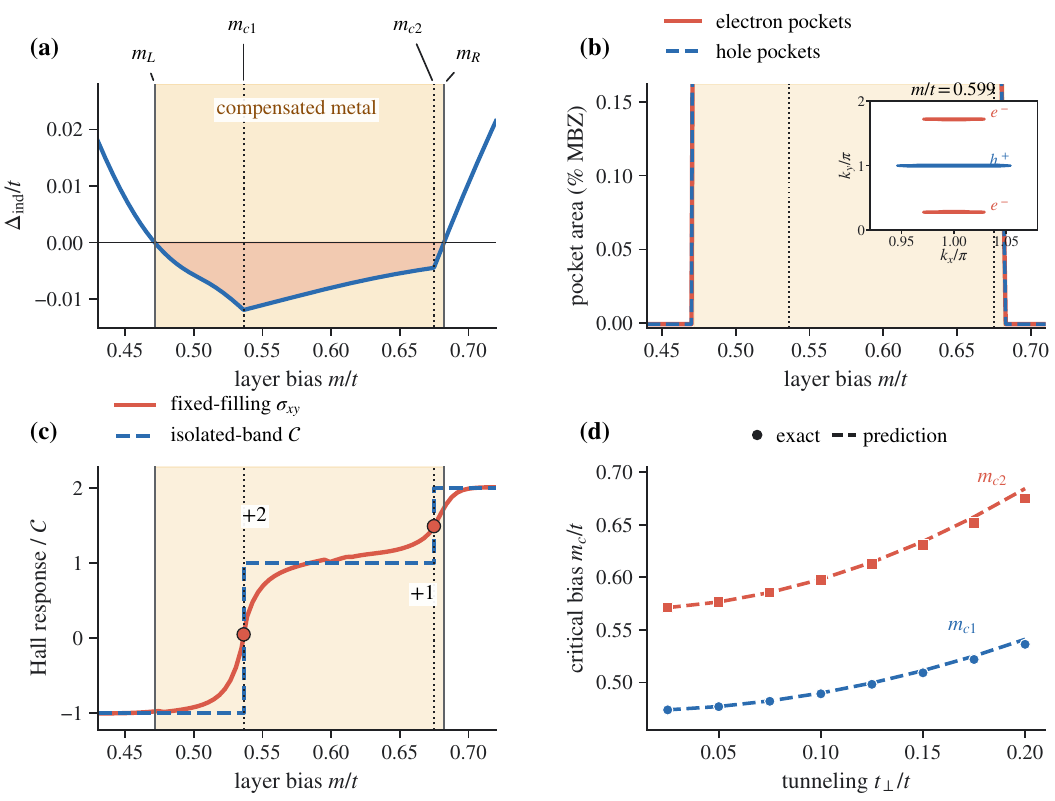}

\caption{\textbf{Topological reconstruction inside a compensated metal.}
(a) The indirect gap is negative only for $m_L<m<m_R$ (shaded).  Dotted lines mark the internal Weyl events $m_{c1},m_{c2}$.
(b) Equal electron and hole pocket areas demonstrate compensation; the inset shows the one hole and two electron Fermi contours at $m/t=0.599$. 
(c) The directly isolated lowest band changes as $-1\to1\to2$, whereas the occupied-state Hall conductivity remains continuous and noninteger.  Independent closed boxes give node charges $(+1,+1)$ at $m_{c1}$ and $+1$ at $m_{c2}$.  Thus Lifshitz and Chern-transfer transitions occur at four distinct biases.(d) Exact critical biases from the full bilayer Hamiltonian (symbols)
compared with the decoupled-layer overlap prediction including the
second-order diagonal energy correction (dashed curves).}
\label{fig:metal}
\end{figure*}

\textit{Generality and channel application.---}
We tested the arithmetic endpoint rule, Eq.~\eqref{eq:endpoint-rule}, against direct diagonalization of the coupled bilayer for 52 rational-flux cases.  In every case the TKNN prediction $\Delta C_{\rm TKNN}=C_2^{\rm TKNN}-C_1^{\rm TKNN}$ agrees with the independently computed endpoint difference $\Delta C_{\rm ED}=C(m_+)-C(m_-)$.  The coarse bias scans separately identify sampled intermediate plateaus, but we do not interpret their telescoping sum as an independent charge test or claim that every narrowly separated transition has been resolved.  In the largest-transfer case, $(2/9,7/9)$, the full Hamiltonian changes from $C=4$ to $C=-4$, while independent closed-surface integrations give eight monopoles of charge $-1$.  A denser arithmetic endpoint survey through $q_i\le30$ contains 13,664 compatible ordered pairs, including 9,776 unequal-denominator cases, and reaches $|\Delta C|=28$~\cite{SM}.

\begin{figure*}[t]
\centering
\includegraphics[width=\textwidth]{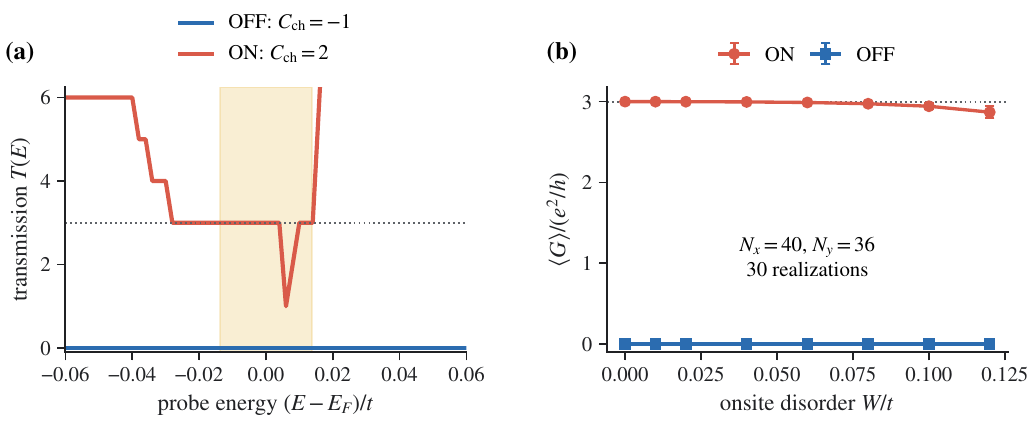}
\caption{ 
(a) Clean transmission of the interface cylinder: the OFF state is insulating, while the ON state has a three-channel plateau.
(b) Disorder-averaged conductance at the common gap center; error bars are one standard deviation over 30 realizations for $W>0$. }
\label{fig:survey}
\end{figure*}

As an example of an electronic application, patterned top and bottom
gates can define a \emph{lateral} channel in the plane of the bilayer.
The differential gate voltage makes the same interlayer bias used
throughout this work spatially dependent,
$m\rightarrow m(\bm r)$, while a common-mode gate can be used to maintain
the desired carrier filling.  In an electronic bilayer,
\begin{equation}
m(\bm r)=U_1(\bm r)-U_2(\bm r)
=-e[\phi_1(\bm r)-\phi_2(\bm r)],
\label{eq:gate-bias}
\end{equation}
where $U_\ell$ and $\phi_\ell$ are the local electrostatic energy and
potential of layer $\ell$, respectively.  A strip-shaped channel can
therefore be set to $m=m_{\rm ch}$ while the surrounding reference
region is held at $m=m_{\rm ref}$.  Their lateral interface carries
\begin{equation}
N_{\rm ch}=|C_{\rm ch}-C_{\rm ref}|
\end{equation}
net chiral modes.  
Differential dual-gate control tunes $m$, while a common-mode voltage can maintain the selected filling. The net interface multiplicity is
\begin{equation}
N_{\rm ch}=|C_{\rm ch}-C_{\rm ref}|.
\end{equation}
The settings $m/t=-1.20$ and $-0.88$ give
$C_{\rm ch}=-1$ and $2$, respectively, and hence
$N_{\rm ch}=0$ and $3$.  We calculate the energy-resolved Landauer
transmission using recursive Green functions,
\begin{equation}
T(E)=\operatorname{Tr}
\!\left[
\Gamma_L(E)G^r(E)\Gamma_R(E)G^a(E)
\right],
\label{eq:transmission}
\end{equation}
where $G^{r,a}$ are the retarded and advanced Green functions of the
scattering region and $\Gamma_{L,R}$ describe its coupling to the leads.
For the spinless model used here, the zero-temperature two-terminal
conductance is
\begin{equation}
G(E_F)=\frac{e^2}{h}T(E_F).
\end{equation}
On a $40\times36$ periodic-transverse cylinder, the common-gap-center
transmissions are
$T_{\rm OFF}<2\times10^{-18}$ and
$T_{\rm ON}=2.99986$, corresponding to
$G_{\rm OFF}<2\times10^{-18}e^2/h$ and
$G_{\rm ON}=2.99986e^2/h$.
Thirty disorder realizations give
$G_{\rm ON}=(2.974\pm0.019)e^2/h$ at $W=0.08t$ and
$(2.943\pm0.034)e^2/h$ at $W=0.10t$
[Fig.~\ref{fig:survey}]. This is a gate-programmable chiral-channel switch, not a voltage gain; its practical OFF leakage and switching speed remain controlled by contacts, disorder, electrostatics, and the unavoidable gapless critical region.

\textit{Implementations.---} The most direct approach with two coupled identical Moir\'e  superlattices would require an unrealistically large gradient of magnetic field across the nanometer separation. 

A more realistic implementation could be two coupled Moir\'e superlattices with different unit cells, made of twisted 2D materials and embedded in a uniform magnetic field. 
Different unit cell areas provide the required magnetic flux mismatch. An experimentally motivated candidate is an asymmetrically aligned hBN--graphene heterostructure in which the two graphene sheets share a common crystallographic orientation while the outer hBN crystals have different orientations and generate two layer-dominant moir\'e potentials with different unit-cell areas; see Fig.~6.  

For the graphene--hBN mismatch $\delta\simeq0.018$, choosing $\theta_b\simeq0$ and $\theta_t\simeq1.02^\circ$ gives $L_b\simeq13.9\,{\rm nm}$, $L_t\simeq9.84\,{\rm nm}$, and $A_b/A_t\simeq2$; at $B\simeq9.87\,{\rm T}$ the corresponding dimensionless fluxes are approximately $(1/5,2/5)$.  These fractions reproduce the magnetic-area ratio of the Harper example but do not by themselves assign its endpoint Chern labels to graphene.  The graphene miniband invariants $C_A^{\rm G}$ and $C_B^{\rm G}$ must be obtained from the material-specific continuum Hamiltonian or from its Hofstadter gap trajectories~\cite{Dean2013}.  The bundle obstruction then enforces $\Delta C_{\rm G}=C_B^{\rm G}-C_A^{\rm G}$.  Only if the relevant graphene endpoint gaps have labels $(-1,2)$, and remain open under the deformation from the Harper model, does the predicted transfer equal $+3$.  Momentum- and sublattice-dependent tunneling may split or relocate the corresponding nodes without changing this conditionally established endpoint difference.  Directly adjacent graphene is a strongly coupled bilayer and is not quantitatively described by $\tp/t=0.2$; a thin hBN spacer is closer to the weak-coupling model.  For the illustrative conversion $t=6\,{\rm meV}$, the closest two events are separated by only $m_R-m_{c2}=0.0436\,{\rm meV}$, so resolving all four transitions requires temperatures and disorder broadening appreciably below this scale.  Coulomb interactions may additionally renormalize or preempt the single-particle cascade.  The quoted biases are therefore design targets pending a material-specific interacting continuum calculation; further implementation constraints are given in Ref.~\cite{SM}.  Projection onto the two magnetic miniband multiplets therefore generates a momentum-dependent matrix $\mathcal T(\mathbf k)$, which can split and relocate the constituent Weyl transitions while preserving the net $(\Delta C=3)$ transfer.

Signature of the metallic state can be revealed with  scanning-SET measurements that resolves incompressible Hofstadter and Chern states in graphene moir\'{e} devices

Other solid-state routes can use effective layer-, valley-, orbital-, strain-, or proximity-dependent fluxes, instead of real magnetic field.  The obstruction also applies directly to two zero-field Chern layers with unequal invariants. 

In anomalous QH effect without any external magnetic field, Berry curvature mismatch will result in a similar cascade of transitions, as sketched in Fig.~7. This scenario which might  even explain the results of the experiment in \cite{Ding2025}. 

Independent fluxes and local intercomponent tunneling are also directly realized in internal-state, synthetic-dimension, photonic, or superconducting-circuit lattices~\cite{Celi2014,Mancini2015,Hafezi2013,Hung2021,Feng2023}. In these settings $m$ is a patterned detuning.  Implementation constraints and a consistent circuit-scale parameter set are detailed in Ref.~\cite{SM}.
\begin{figure*}[t]
\includegraphics[width=\textwidth]{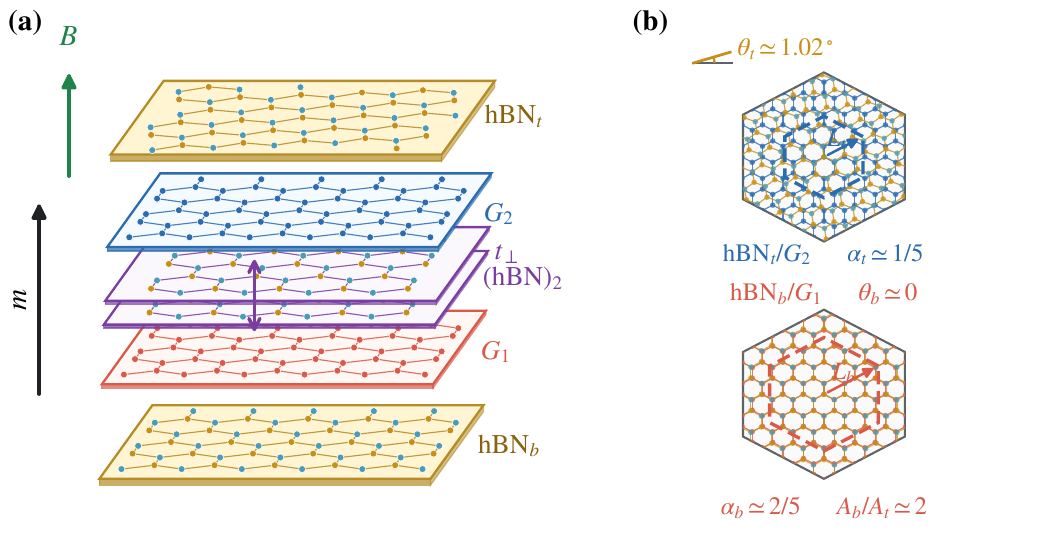}
\caption{\textbf{Asymmetric hBN--graphene realization.}
(a) Aligned graphene sheets separated by a two-layer hBN spacer and encapsulated by differently oriented hBN.  The spacer controls $t_\perp$, while $m$ and $B$ tune the bias and magnetic flux.
(b) Overlaid interface lattices.  The choices $\theta_b\simeq0$ and $\theta_t\simeq1.02^\circ$ give $A_b/A_t\simeq2$ and representative fluxes $(\alpha_t,\alpha_b)\simeq(1/5,2/5)$.  The displayed twist is enlarged for visibility.}
\label{fig:realization_hbn}
\end{figure*}

\begin{figure*}[t]
\includegraphics[width=\textwidth]{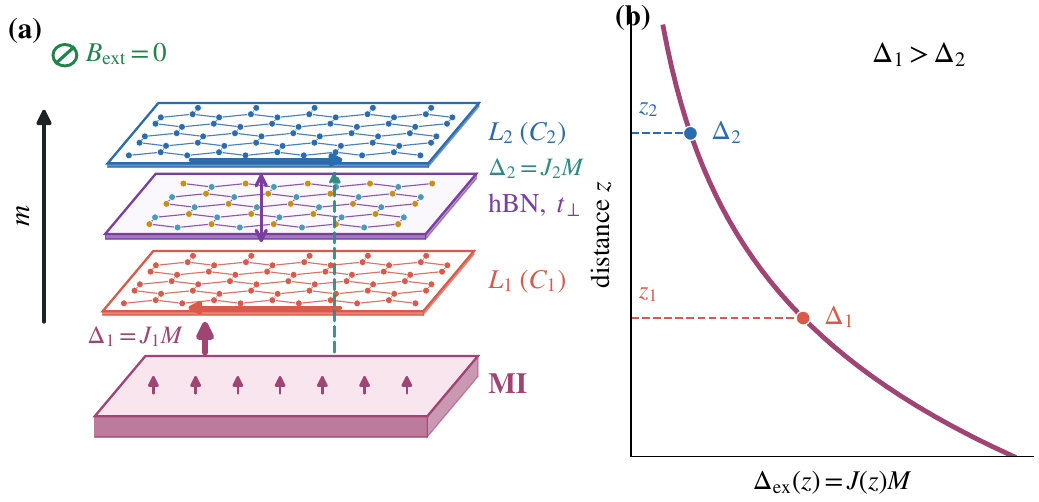}
\caption{\textbf{Single-substrate anomalous-Hall realization.}
(a) Two weakly coupled electronic layers lie at different distances from the same magnetic-insulator (MI) substrate.  The proximity exchange is therefore unequal, $\Delta_i=J_iM$ with $\Delta_1>\Delta_2$, while the layer bias $m$ is electrically tunable and $B_{\rm ext}=0$.
(b) Schematic decay of the exchange field away from the substrate.  Different exchange gaps can produce isolated anomalous-Hall bands with unequal invariants $C_1\ne C_2$, to which the Chern-transfer mechanism applies.}
\label{fig:realization_qah}
\end{figure*}

\textit{Conclusion.---}
We proposed a robust and experimentally accessible  mechanism of electric control of the  topological electronic states. The underlying physics is flux mismatch between the two layers which converts an ordinary band inversion into a quantized transfer of Berry charge.  Two possible routes of Chern transfer are identified. The proposed implementation of electrically programmable Chern phases and their chiral signatures can be realized in a variety of solid-sate and synthetic platforms, most  straightforwardly in stacked 2D materials.   

\IfFileExists{apsrev4-2.bst}{\bibliographystyle{apsrev4-2}}{\bibliographystyle{unsrt}}
\ifapsclass\else\footnotesize\fi
\bibliography{references}

\begin{thebibliography}{10}

\bibitem{Hofstadter1976}
Douglas~R. Hofstadter.
\newblock Energy levels and wave functions of bloch electrons in rational and irrational magnetic fields.
\newblock {\em Phys. Rev. B}, 14:2239--2249, 1976.

\bibitem{Dean2013}
C.~R. Dean, L.~Wang, P.~Maher, C.~Forsythe, F.~Ghahari, Y.~Gao, J.~Katoch, M.~Ishigami, P.~Moon, M.~Koshino, T.~Taniguchi, K.~Watanabe, K.~L. Shepard, J.~Hone, and P.~Kim.
\newblock Hofstadter's butterfly and the fractal quantum hall effect in moir\'e superlattices.
\newblock {\em Nature}, 497:598--602, 2013.

\bibitem{Hunt2013}
B.~Hunt, J.~D. Sanchez-Yamagishi, A.~F. Young, M.~Yankowitz, B.~J. LeRoy, K.~Watanabe, T.~Taniguchi, P.~Moon, M.~Koshino, P.~Jarillo-Herrero, and R.~C. Ashoori.
\newblock Massive dirac fermions and hofstadter butterfly in a van der waals heterostructure.
\newblock {\em Science}, 340:1427--1430, 2013.

\bibitem{Ponomarenko2013}
L.~A. Ponomarenko, R.~V. Gorbachev, G.~L. Yu, D.~C. Elias, R.~Jalil, A.~A. Patel, A.~Mishchenko, A.~S. Mayorov, C.~R. Woods, J.~R. Wallbank, M.~Mucha-Kruczy\'nski, B.~A. Piot, M.~Potemski, I.~V. Grigorieva, K.~S. Novoselov, F.~Guinea, V.~I. Fal'ko, and A.~K. Geim.
\newblock Cloning of dirac fermions in graphene superlattices.
\newblock {\em Nature}, 497:594--597, 2013.

\bibitem{Spanton2018}
Eric~M. Spanton, A.~A. Zibrov, H.~Zhou, T.~Taniguchi, K.~Watanabe, M.~P. Zaletel, and A.~F. Young.
\newblock Observation of fractional chern insulators in a van der waals heterostructure.
\newblock {\em Science}, 360:62--66, 2018.

\bibitem{Chen2020Chern}
Guorui Chen, Aaron~L. Sharpe, Eli~J. Fox, Ya-Hui Zhang, Shiming Wang, Lili Jiang, Bo~Lyu, Hongyuan Li, Kenji Watanabe, Takashi Taniguchi, Zhiwen Shi, T.~Senthil, David Goldhaber-Gordon, and Yuanbo Zhang.
\newblock Tunable correlated chern insulator and ferromagnetism in a moir\'e superlattice.
\newblock {\em Nature}, 579:56--61, 2020.

\bibitem{Ding2025}
Jing Ding, Hanxiao Xiang, Jiannan Hua, Wenqiang Zhou, Naitian Liu, Le~Zhang, Na~Xin, Bing Wu, Kenji Watanabe, Takashi Taniguchi, Zden\v{e}k Sofer, Wei Zhu, and Shuigang Xu.
\newblock Electric-field switchable chirality in rhombohedral graphene chern insulators stabilized by tungsten diselenide.
\newblock {\em Phys. Rev. X}, 15:011052, 2025.

\bibitem{Perrin2024}
Mickael~L. Perrin, Anooja Jayaraj, Bhaskar Ghawri, Kenji Watanabe, Takashi Taniguchi, Daniele Passerone, Michel Calame, and Jian Zhang.
\newblock Electric field tunable bandgap in twisted double trilayer graphene.
\newblock {\em npj 2D Mater. Appl.}, 8:14, 2024.

\bibitem{Patra2023}
Sumanti Patra, Prasun Boyal, and Priya Mahadevan.
\newblock Electric-field-induced metal-semiconductor transitions in twisted bilayers of {WSe$_2$}.
\newblock {\em Phys. Rev. B}, 107:L041104, 2023.

\bibitem{Huang2016}
Zhoushen Huang and Alexander~V. Balatsky.
\newblock Dynamical quantum phase transitions: Role of topological nodes in wave function overlaps.
\newblock {\em Phys. Rev. Lett.}, 117:086802, 2016.

\bibitem{Bultinck2020}
Nick Bultinck, Shubhayu Chatterjee, and Michael~P. Zaletel.
\newblock Mechanism for anomalous hall ferromagnetism in twisted bilayer graphene.
\newblock {\em Phys. Rev. Lett.}, 124:166601, 2020.

\bibitem{Hu2015ChernMetal}
Xiang Hu, Zhicheng Zhong, and Gregory~A. Fiete.
\newblock First principles prediction of topological phases in thin films of pyrochlore iridates.
\newblock {\em Sci. Rep.}, 5:11072, 2015.

\bibitem{Sorn2018}
Sopheak Sorn.
\newblock Bilayer {Haldane} model: From trivial band insulator to fractionalized quantum anomalous hall insulator.
\newblock {\em Phys. Rev. B}, 98:125145, 2018.

\bibitem{TKNN1982}
D.~J. Thouless, M.~Kohmoto, M.~P. Nightingale, and M.~den Nijs.
\newblock Quantized hall conductance in a two-dimensional periodic potential.
\newblock {\em Phys. Rev. Lett.}, 49:405--408, 1982.

\bibitem{Wannier1978}
G.~H. Wannier.
\newblock A result not dependent on rationality for bloch electrons in a magnetic field.
\newblock {\em Phys. Status Solidi B}, 88:757--765, 1978.

\bibitem{DanaAvronZak1985}
I.~Dana, Y.~Avron, and J.~Zak.
\newblock Quantised hall conductance in a perfect crystal.
\newblock {\em J. Phys. C: Solid State Phys.}, 18:L679--L683, 1985.

\bibitem{SM}
See Supplemental Material for magnetic Bloch conventions, the strong-tunneling expansion, the bundle proof and multiband extension, the systematic rational-flux survey, recursive Green-function transport, additional critical-point data, and implementation details.

\bibitem{Fukui2005}
Takahiro Fukui, Yasuhiro Hatsugai, and Hiroshi Suzuki.
\newblock Chern numbers in discretized brillouin zone: Efficient method of computing (spin) hall conductances.
\newblock {\em J. Phys. Soc. Jpn.}, 74:1674--1677, 2005.

\bibitem{Schnyder2008}
Andreas~P. Schnyder, Shinsei Ryu, Akira Furusaki, and Andreas W.~W. Ludwig.
\newblock Classification of topological insulators and superconductors in three spatial dimensions.
\newblock {\em Phys. Rev. B}, 78:195125, 2008.

\bibitem{Ryu2010}
Shinsei Ryu, Andreas~P. Schnyder, Akira Furusaki, and Andreas W.~W. Ludwig.
\newblock Topological insulators and superconductors: tenfold way and dimensional hierarchy.
\newblock {\em New J. Phys.}, 12:065010, 2010.

\bibitem{Celi2014}
Alessio Celi, Pietro Massignan, Julius Ruseckas, Nathan Goldman, Ian~B. Spielman, Gediminas Juzeli\=unas, and Maciej Lewenstein.
\newblock Synthetic gauge fields in synthetic dimensions.
\newblock {\em Phys. Rev. Lett.}, 112:043001, 2014.

\bibitem{Mancini2015}
Marco Mancini, Giacomo Pagano, Giacomo Cappellini, Leonardo Livi, Marie Rider, Jacopo Catani, Carlo Sias, Peter Zoller, Massimo Inguscio, Marcello Dalmonte, and Leonardo Fallani.
\newblock Observation of chiral edge states with neutral fermions in synthetic hall ribbons.
\newblock {\em Science}, 349:1510--1513, 2015.

\bibitem{Hafezi2013}
Mohammad Hafezi, S.~Mittal, J.~Fan, A.~Migdall, and J.~M. Taylor.
\newblock Imaging topological edge states in silicon photonics.
\newblock {\em Nat. Photon.}, 7:1001--1005, 2013.

\bibitem{Hung2021}
Jimmy S.~C. Hung, J.~H. Busnaina, C.~W.~Sandbo Chang, A.~M. Vadiraj, I.~Nsanzineza, E.~Solano, H.~Alaeian, E.~Rico, and C.~M. Wilson.
\newblock Quantum simulation of the bosonic {Creutz} ladder with a parametric cavity.
\newblock {\em Phys. Rev. Lett.}, 127:100503, 2021.

\bibitem{Feng2023}
Wei Feng, Dexi Shao, Guo-Qiang Zhang, Qi-Ping Su, Jun-Xiang Zhang, and Chui-Ping Yang.
\newblock Quantum simulation of {Hofstadter} butterfly with synthetic gauge fields on two-dimensional superconducting-qubit lattices.
\newblock {\em Front. Phys.}, 18:61302, 2023.

\end{thebibliography}

\end{document}